# Sergey Porotsky

# Is it Correct to Use MLE Method for GRP Parameter Estimation

*Abstract* - Analysis of repair systems usually uses an ''as good as new'' or ''as bad as old'' repair assumptions. In practice, repair actions do not result in such extreme situations, but rather in a complex transitional one, that is imperfect maintenance, i.e. Generalized Renewal Process (GRP). Maximum Likelihood Estimation method is often used for reliability parameter estimation, but is it correct to use it for GRP ?

## I. INTRODUCTION

Maintenance action is carried out either to preserve a system or to renovate it to a special state. This maintenance can be divided into corrective maintenance (CM), which is carried out after failure, and preventive maintenance (PM), that is carried out after a certain time interval or age without any failure. PM is carried out when the system is operating, and intends to slow down the wear process to reduce the frequency of occurrence of system failures. PM can be time-based or condition-based. Condition-based PM occurs at unscheduled times, which are determined according to the results of inspections, and degradation or operation controls.

Maintenance action is also characterized by the degree to which a system can be restored. The most common assumptions on maintenance efficiency are known as minimal repair, or "As Bad As Old" (ABAO), and perfect repair, or "As Good As New" (AGAN). In the ABAO case, each maintenance leaves the system in the state it was before maintenance. In the AGAN case, each maintenance is perfect, and leaves the system as if it were new. It is common for true experience to result between these behaviors: standard maintenance reduces failure intensity, but does not leave the system as good as new. This situation is known as imperfect maintenance.

A lot of models have been developed for imperfect repairs that assume that component is "better than old but worse than new" after repair. The approach of Brown and Proschan [1] considered that the anticipated repair is either a perfect repair with probability p or a minimal repair with probability (1- p). One of the most popular is the Generalized Renewal Process (GRP), introduced by Kijima and Sumita [2]. They established an effectiveness parameter q, defining a virtual age of the system component at a given time after several repairs:

$$V_i = qS_i \qquad (1)$$

where $V_i$ is virtual age, $S_i$ is real time immediately after i-th repair and q is restoration factor. If $q = 0$, the virtual age of a unit is set to zero, which corresponds to the "As Good As New" repair assumption and represents the ordinary renewal process. If $q = 1$, the component is considered "As Bad As Old" after restoration, which is the case of the Non-Homogeneous Poisson Process (NHPP).

Equation (1) is proved only for situation, when value of the parameter q isn't changed during item life. If value of the q can change (i.e., it may be different for CM and PM), other equations should be used:

$$V_i = V_{i-1} + q_i t_i \qquad (2)$$

In this equation the $t_i$ is the length between (i-1) and i-th events, $V_i$ is the virtual age of the item after the i-th maintenance action, $q_i$ is the restoration factor of imperfect maintenance after i-th event ($0 \leq q_i \leq 1$). The effect of the i-th maintenance is to reduce the virtual age just before the moment of failure, by an amount proportional to the time elapsed since the previous maintenance.

We will consider the most popular Weibull failure time CDF, which is

$$F(t) = 1 - \exp(-a(t^b)) \qquad (3)$$

where a is scale parameter and b is shape parameter (sometimes as scale parameter is used parameter teta = $(1/a)^{(1/b)}$ ). To estimate the reliability parameters (a and b) and maintainability parameters ($q_{pm}$ for PM and $q_{cm}$ for CM) the Maximum Likelihood Estimation (MLE) method is often used. Currently the approach of Yanez et al. [3] is among the most widely used for GRP parameter estimation. The inter-arrival of failures are assumed to follow the Weibull distribution, times for PM and CM are assumed as negligible and so the likelihood functions for i-th event are given by the following expressions:

$L_i = \exp(-a((V_{i-1} + t_i)^b - V_{i-1}^b))$ , if i-th event is PM

$L_i = (ab((V_{i-1} + t_i)^{(b-1)}))\exp(-a((V_{i-1} + t_i)^b - V_{i-1}^b))$ , if i-th event is CM

$V_i$ are calculated according equation (2) and $V_0 = 0$.

Full Likelihood (for all events) is multiplication of $L_i$.

## II. RESULTS of RESEARCH

### 2.1. Initial Estimations

To estimate parameters a (or teta), b, $q_{pm}$ and $q_{cm}$ the Cross-Entropy optimization method [4] was used. Really instead of Likelihood maximization the LogLikelihood was maximized. To test our MLE-based tools for GRP parameter estimation first we have used data set, proposed on [5]. On [5] were generated 100 points according following initial values of parameters:

***teta = 1, b = 2.2, $q_{pm}$ = 0.8, $q_{cm}$ = 0.3***

Unfortunately, our results were very far from initial values of restoration parameters of $q_{pm}$ and $q_{cm}$, used for generation. We have got following values:

***teta = 1.12, b = 2.54, $q_{pm}$ = 1.0, $q_{cm}$ = 0.0***

It is necessary to note, that very large difference were observed only for restoration parameters, for scale and shape parameter difference are not large (12 % and 15%).

It is also necessary to note, that our results were very far not only from initial values of parameters, but also from values, which were got by means of other method using. To estimate parameters teta, b, $q_{pm}$ and $q_{cm}$ on the [5] instead of MLE, the Bayesian/Sampling Method was used.

The following mean values were obtained on [5]:

*teta = 1.03, b = 2.34, q$_{pm}$ = 0.74, q$_{cm}$ = 0.47*

Generally speaking, difference between our values and both initial values and values from [5] are not "rare event". Different methods (e.g., Moments, Least Squares, MLE) sometimes can get essentially different values of estimated parameters. Sometimes, randomly, difference between initial parameters and parameters (used for sample generation), and values, received by means of MLE, may be large.

So, we have searched and discovered results of other estimations of the data set, generated on [5]. Articles [6, 7] also have used this statistics for GRP parameter estimations, based on MLE method. What is strange, that results are very far from our and very similar for mentioned on [5]. On [6] the following values were obtained:

*teta = 0.997, b = 2.26, q$_{pm}$ = 0.73, q$_{cm}$ = 0.45*

On the [7] Genetic Algorithm (GA) was used for MLE optimization and were performed 20 realizations of the Maximum LogLikelihood searchs. Mean values were following:

*teta = 1.09, b = 2.20, q$_{pm}$ = 0.82, q$_{cm}$ = 0.38*

and values, corresponding for Maximum (for all 20 realizations) of the Maximum LogLikelihood, were following:

*teta = 1.09, b = 2.24, q$_{pm}$ = 0.82, q$_{cm}$ = 0.33*

To compare all results more detail, we have calculated LogLikelihood for them. Results are summarized below:

| Source | Method | teta | b | q$_{pm}$ | q$_{cm}$ | LogLikelihood Value |
|---|---|---|---|---|---|---|
| [5] | Initial (for sample generation) | 1.0 | 2.2 | 0.8 | 0.3 | 58.32 |
| [5] | Bayesian/Sampling | 1.03 | 2.34 | 0.74 | 0.47 | 55.7 |
| [6] | MLE | 0.997 | 2.26 | 0.73 | 0.44 | 57.01 |
| [7] | MLE and GA, mean values | 1.09 | 2.20 | 0.82 | 0.38 | 58.07 |
| [7] | MLE and GA, best values | 1.09 | 2.24 | 0.82 | 0.33 | 58.12 |
| Our | MLE | 1.12 | 2.54 | 1.0 | 0.0 | 58.96 |

On [6] also mentioned results of LogLikelihood calculations for some values of

{teta, b, q$_{pm}$ and q$_{cm}$} and they very differ from our calculations. For example, for set

{teta = 1, b = 2.21, q$_{pm}$ = 0.73 and q$_{cm}$ = 0.44 } according [6] LogLikelihood = 68.04, but our calculations according expression (10) from [6] have got result LogLikelihood = 57.84

So, or we have calculated LogLikelihood incorrect, or on [6] LogLikelihood was calculated incorrect.

Analyzing data from above table, we can conclude:

- It isn't strange, that initial values (second row) and values, obtained by means of other method (third row) have more low Likelihood values, that our values (last row).

- What is really strange, that values of rows 4, 5 and 6, obtained also by means of MLE method, have more low values of Likelihood, that our values and so really they are not MLE solutions!

  So, we can formulate following reasons for this large difference between our value and initial values, used for generation:

- It was error (some bug) on the our implementation of the Likelihood calculation for GRP with both PM and CM

- It was error (some bug) on the our implementation of the Likelihood maximization

- It was randomly, only for this concrete sample

- It is typical situation for MLE using for GRP parameter estimation

To search real reason, we have performed additional research.

## 2.2 Analysis of Estimations, based on Other Statistics

First we have discovered some articles, which have solved same task (GRP parameter estimation by means of MLE), but based on other statistics.

On [8] the following task is solved: to estimate parameters of the GRP, which consists of only CM. Input statistics is times of 24 failures and following values were got on [8]:

***a = 0.00494, b = 1.198, q = 0.1344, Maximum LogLikelihood Value = -123.6347***

Using of our tools, we have got following values:

***a = 0.00489, b = 1.200, q = 0.1260, Maximum LogLikelihood Value = -123.6353***

So, for this statistics our results and results of ReliaSoft are same.

On [9] the following task is solved : to estimate parameters of the GRP, which consists of both CM and PM. Input statistics are times of 11events ( 4 PM and 7 CM) and following values were got on [9]:

***a = 1.76e-4, b = 2.36, $q_{cm}$ = 0.06, $q_{pm}$ = 0.0***

Using of our tools, we have got following values:

***a = 1.72e-4, b = 2.3652, $q_{cm}$ = 0.0602, $q_{pm}$ = 0.00015***

So, for this statistics our results and results of [9] are same.

*Same results of estimations for these two tasks allow us to conclude, that our tools perform both Likelihood Calculation and Likelihood Maximization fully correct. So, there were no error on the our estimations on chapter 2.1 and rather were errors on the optimization on the [6] and [7] – it values don't correspond for Maximum Likelihood.*

## 2.3 Detailed Analysis of Estimations, based on Initial Statistics

Return to the question, formulated above, on the Chapter 2.1 – what is reason for the large difference between initial parameters (used for sample generation), and our values, received by means of MLE. Now we are sure, that first and second reasons are not true, so may be only two reasons:

- It was randomly, only for this concrete sample
- It is typical situation for MLE using for GRP parameter estimation

First we have wanted to check stability of the results. Based on MLE using. We could not change any intermediate times from initial sample of 100 values, but we can change last value. We see, that from restoration factor point of view, *the stability isn't observed:*

For value $t_{100} = 0.35$ we have got following MLE-based estimations:

***teta = 0.94, b = 2.0, $q_{pm}$ = 1.0, $q_{cm}$ = 0.35,***

and for value $t_{100} = 0.45$ we have got following estimations:

***teta = 0.95, b = 1.8, $q_{pm}$ = 1.0, $q_{cm}$ = 0.94.***

To check more detail, that this large difference between initial parameters (used for sample generation) and our values isn't random, i.e. not only for this current sample from [5], we have generated additional 20 samples of 100 times and have estimated reliability and restoration parameters for them. Unfortunately, results are same, that described at the Chapter 2.1 – on the most samples (14 from 20) both restoration factors $q_{pm}$ and $q_{cm}$ have got extreme values, 0 or 1 and error is too large. Results of estimations are below:

| Sample Number | b | teta | qpm | qcm |
|---|---|---|---|---|
| 1 | 2.6 | 1.6 | 1.0 | 0.2 |
| 2 | 2.2 | 1.1 | 1.0 | 0.0 |
| 3 | 2.9 | 1.8 | 1.0 | 0.0 |
| 4 | 3.0 | 1.4 | 1.0 | 0.0 |
| 5 | 3.0 | 1.5 | 0.8 | 0.0 |
| 6 | 2.0 | 1.2 | 1.0 | 1.0 |
| 7 | 2.5 | 1.2 | 1.0 | 0.0 |
| 8 | 2.2 | 0.7 | 0.4 | 0.0 |
| 9 | 2.6 | 1.4 | 1.0 | 0.1 |
| 10 | 2.8 | 1.4 | 1.0 | 0.0 |
| 11 | 2.0 | 1.2 | 1.0 | 1.0 |
| 12 | 2.4 | 1.3 | 1.0 | 0.0 |
| 13 | 2.2 | 1.0 | 1.0 | 0.2 |
| 14 | 2.8 | 0.8 | 0.3 | 0.1 |
| 15 | 2.9 | 1.6 | 1.0 | 0.0 |
| 16 | 2.9 | 0.8 | 0.2 | 0.1 |
| 17 | 3.0 | 0.9 | 0.8 | 0.0 |
| 18 | 2.7 | 0.8 | 0.0 | 0.4 |
| 19 | 2.1 | 1.2 | 1.0 | 1.0 |
| 20 | 2.5 | 0.8 | 0.7 | 0.0 |

## 2.4. Advanced Analysis of Estimations, based on Initial Statistics

To generate initial statistics of the 100 PM and CM, on the [5] the simplest strategy was used – independently, according conditional Weibull distribution, were generated $t_{cm}$ and $t_{pm}$. If $t_{cm} < t_{pm}$, the next event will be CM, else next event will be PM. Same rule we have used on the chapter 2.3 to generate 20 samples of 100 CM and PM.

According this rule, amount of generated CM and PM events approximately equaled. In real life amount of PM in some systems may be essentially more, that amount of the CM. To take into account these possible situations, we will use more complex rule. Insert additional parameter $K_{cm}$, which characterize frequency of the CM. Updated rule to generate next event is following:

$t_{cm} = ( V^b - \log(1- rand)/a )^{(1/b)} - V$; $t_{pm} = K_{cm}*( ( V^b - \log(1- rand)/a )^{(1/b)} – V )$;

if $t_{cm} <  t_{pm}$, the next event will be CM else else next event will be PM.

We were generated 20 samples of the 100 PM and CM according $K_{cm} = 0.1$ (it was approximately 10 CM and 90 PM on the single sample) and 20 samples of the 100 PM and CM according $K_{cm} = 0.3$ (it was approximately 20 CM and 80 PM on the single sample). Results are summarized on the table below. As in previously situation, described on the chapter 2.3 (when $K_{cm} = 1.0$), on the most samples the restoration factors $q_{pm}$ and $q_{cm}$ have got extreme values, 0 or 1.

| Sample Number | Kcm = 0.1 | | Kcm = 0.3 | |
|---|---|---|---|---|
| | $q_{pm}$ | $q_{cm}$ | $q_{pm}$ | $q_{cm}$ |
| 1 | 1.0 | 0.0 | 1.0 | 0.0 |
| 2 | 1.0 | 0.0 | 1.0 | 0.0 |
| 3 | 0.4 | 0.0 | 1.0 | 0.0 |
| 4 | 1.0 | 0.0 | 0.6 | 0.0 |
| 5 | 1.0 | 0.0 | 0.5 | 0.0 |
| 6 | 1.0 | 0.0 | 1.0 | 0.0 |
| 7 | 0.3 | 0.0 | 1.0 | 0.0 |
| 8 | 1.0 | 0.0 | 1.0 | 0.0 |
| 9 | 0.3 | 0.0 | 0.3 | 0.0 |
| 10 | 1.0 | 0.0 | 1.0 | 0.5 |
| 11 | 1.0 | 0.0 | 0.6 | 0.0 |
| 12 | 1.0 | 0.0 | 1.0 | 0.0 |
| 13 | 0.4 | 0.6 | 1.0 | 0.0 |
| 14 | 1.0 | 0.0 | 0.5 | 0.0 |
| 15 | 0.6 | 0.0 | 0.6 | 0.0 |
| 16 | 0.8 | 0.0 | 0.6 | 0.0 |
| 17 | 1.0 | 0.0 | 0.3 | 0.0 |
| 18 | 1.0 | 0.0 | 1.0 | 0.0 |
| 19 | 1.0 | 0.0 | 0.5 | 0.0 |
| 20 | 0.9 | 0.0 | 0.5 | 0.0 |

## 2.5. Analysis of Estimations, based on Large Statistics

It may be seemed, that very bad accuracy for restoration factor estimations is only due to the sample size (100 events) and for more large size the accuracy will be good. It is impossible directly to generate sample of more large size, e.g. of size 300…1000 events. Reason is following – at the end of the sample aging will be too large and both $t_{cm}$ and $t_{pm}$ will be too small. So, instead of large sample generation we were used a few (3…10) independent items and have estimated required GRP parameters for all items simultaneously. Likelihood for all items will be multiplication of the likelihoods of single items.

Results are summarized on the table below. In differ from previously situation, described on the chapter 2.3 (when sample size was 100 events), now on the most samples restoration factor $q_{cm}$ have not got extreme values (0 or 1). But restoration factor $q_{pm}$, event for size of 1000 events, very often have got extreme value 1 and accuracy yet isn't good.

| Sample Number | 300 PM&CM | | 1000 PM&CM | |
|---|---|---|---|---|
| | qpm | qcm | qpm | qcm |
| 1 | 1.0 | 0.4 | 0.8 | 0.4 |
| 2 | 1.0 | 0.7 | 0.8 | 0.3 |
| 3 | 1.0 | 0.2 | 1.0 | 0.3 |
| 4 | 0.3 | 0.2 | 0.4 | 0.2 |
| 5 | 1.0 | 0.0 | 1.0 | 0.0 |
| 6 | 0.8 | 0.2 | 1.0 | 0.3 |
| 7 | 1.0 | 0.6 | 0.6 | 0.3 |
| 8 | 1.0 | 0.0 | 0.5 | 0.1 |
| 9 | 0.5 | 0.2 | 0.5 | 0.2 |
| 10 | 0.3 | 0.1 | 0.7 | 0.2 |
| 11 | 0.9 | 0.6 | 1.0 | 0.4 |
| 12 | 0.4 | 0.2 | 1.0 | 0.4 |
| 13 | 0.6 | 0.1 | 0.6 | 0.2 |
| 14 | 0.6 | 0.1 | 0.7 | 0.1 |
| 15 | 0.5 | 0.2 | 0.5 | 0.2 |
| 16 | 1.0 | 0.3 | 0.7 | 0.2 |
| 17 | 0.3 | 0.2 | 1.0 | 0.3 |
| 18 | 0.4 | 0.3 | 0.5 | 0.2 |
| 19 | 0.8 | 0.1 | 0.4 | 0.2 |
| 20 | 1.0 | 0.2 | 1.0 | 0.2 |

## III. CONCLUSIONS

During testing of our MLE-based tools for GRP parameter estimations we have discovered following:

- Accuracy of restoration parameter estimation, both for Preventive and Corrective Maintenance, is too low. In most cases estimated values are only extreme, i.e. 0 or 1.

- Information about good accuracy of MLE using for GRP estimation, published on some articles, isn't correct – really these estimations don't correspond for Maximum Likelihood.

- To check possibility to guarantee high accuracy of the received estimations in each concrete task, it is recommended to perform Monte-Carlo simulation by means of generation a few samples with received parameter values.

## Acknowledgment

The author would like to thank Pingjian Yu, one of the article [5] authors, for useful discussion.